\documentstyle[aps,pra,epsfig]{revtex}
\begin{document}

\def\be{\begin{eqnarray}}
\def\ee{\end{eqnarray}}
\title{Localization and diffusion in Ising-type quantum networks}
\author{P. T\"orm\"a$^{(a)}$, I. Jex$^{(b)}$, and W.P.\ Schleich$^{(c)}$}
\date{\today}
\maketitle
\begin{description}
\item [(a)] Laboratory of Computational Engineering, P.O.B. 9400,
FIN-02015 Helsinki University of Technology, Finland \\
\item[(b)] Department of Physics, FNSPE, Czech Technical University Prague,
B\v rehov\'a 7, 115 19 Praha 1, Czech Republic\\
\item [(c)] Abteilung f\"ur Quantenphysik, Universit\"at Ulm,
Albert-Einstein Allee 15, D-89081 Ulm, Germany\\
\end{description}
\begin{abstract} We investigate the effect of phase randomness in Ising-type
quantum networks. These networks model a large class of physical
systems. They describe micro- and
nanostructures or arrays of optical elements such as beam splitters (interferometers)
or parameteric amplifiers. Most of these stuctures are promising candidates for
quantum information processing networks.
We demonstrate that such systems exhibit two
very distinct types of behaviour. For certain network configurations
(parameters), they show quantum localization similar to
Anderson localization whereas
classical stochastic behaviour is observed in other cases.
We relate these findings to the standard theory of quantum localization.
\end{abstract}

\section{Introduction}

Maturing of the techniques for controlling single quantum systems such
as atoms, photons, or quantum dots has made it possible to design more
complicated structures of these basic elements: quantum networks.
Interesting examples of such micro- and nanostructures are for instance thin
wires used as atom guides \cite{Denschlag99}, trapped chains of atoms and
ions \cite{Wineland01}, or structures grown on a substrate such as
quantum dots \cite{Mahler} or arrays of nanomagnets.  These
are among the first steps of laboratory demonstrations leading to
quantum network engineering and finally to matured and scalable quantum
technology. Due to these possible applications investigations of
quantum networks have become topical.

In this paper we consider a quantum network configuration which is perhaps the
simplest non-trivial one: a network which realizes nearest-neighbour
interactions by coupling the individual quantum systems {\it pairwise}.
As a possible implementation one can think of light modes
coupled together by an optical element, for example a beam splitter or
a mode coupler. The network is schematically presented in Fig.1.
Another common form of nearest neighbour interaction is simulataneous
interaction with many neighbours, like electrons interacting with
Coulomb interaction. Here we, however, consider pairwise interaction
which is typical for light modes, and for many quantum networks in general.

\vskip1truecm
\hskip3truecm\includegraphics[width=5cm,height=5cm]{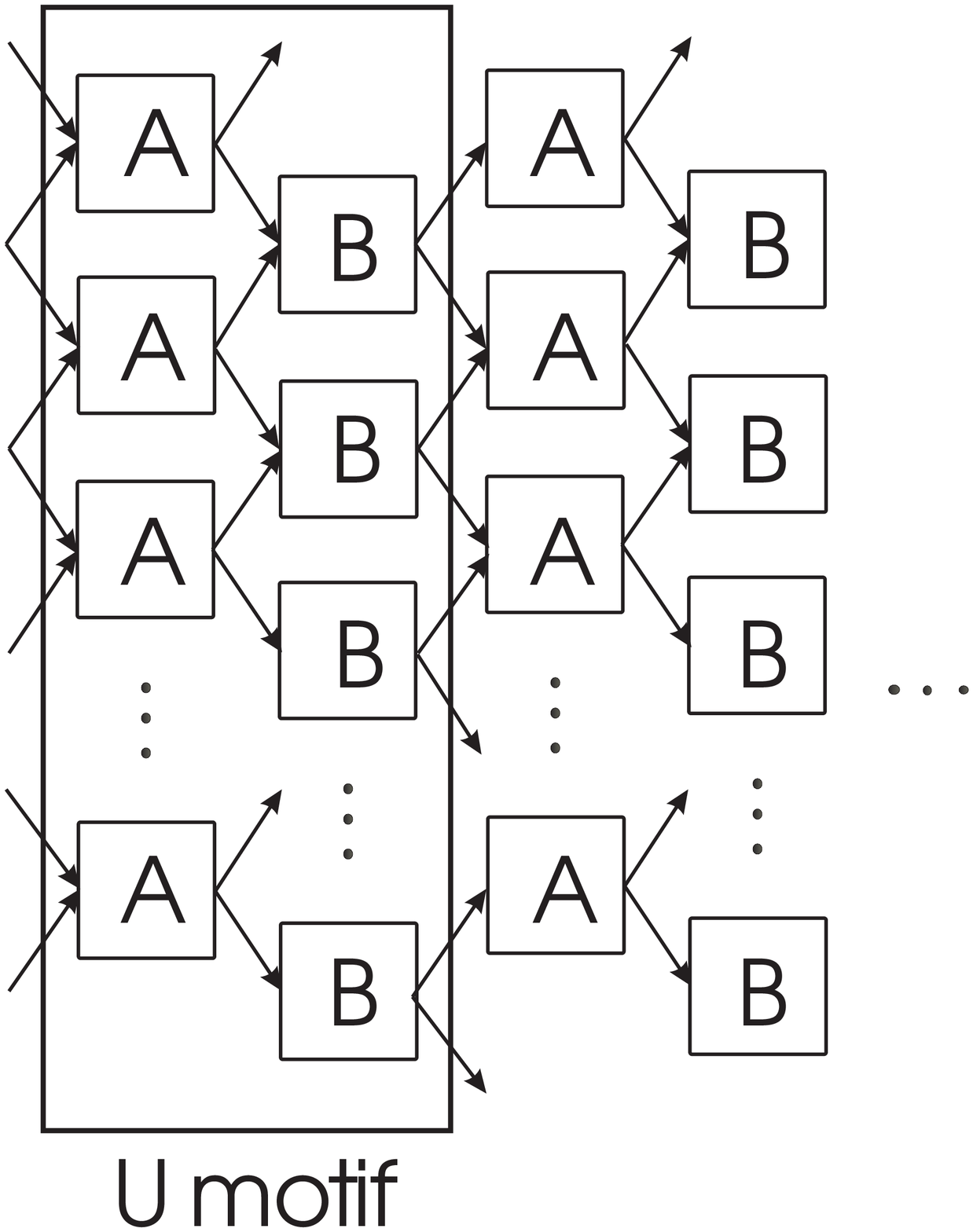}\\

\begin{figure}
\caption{A schematic presentation of the considered quantum network.
The nodes are connected to their
nearest neighbours. The boxes \fbox{A} and \fbox{B} denote the
transformations performed at the nodes; they could correspond for
instance to beam splitters in case of an optical realization.}
\end{figure}

The network in Fig.1 has an interesting and potentially useful
connection to the Ising-model \cite{Torma98,Kraus}.
Interference effects are typical in this kind of networks and have
been investigated by many authors
\cite{TJ1,Bouwmeester,Harmin,others}. The Hamiltonian structure of the
network was studied in \cite{TJ2}.  In this paper we consider the
effect of randomness in such a network. Randomness in certain quantum
systems is know to lead to intriguing effects such as Anderson
localization and dynamical localization
\cite{Anderson58,Haake92,Casati95}. Also classical wave behaviour
combined with randomness can lead to localization, c.f.\ localization
of light \cite{Berry97,Delande00}

We can view the considered network also as a interferometer. The effects
demonstrated can be understood as the result of interferences between paths
available for the excitations in the network. However, interferences are
known to be very fragile towards phase perturbations.
We show that the typical interference effects are modified or
completely destroyed as randomness is introduced in the network. 

The way in which the randomness enters the network has a profound effect
on the behaviour. Therefore we consider two basic variants of the
network.  First, a 2-D network of the type shown in Fig.1, with
randomness all over the network. This means that along each connecting
line between two interaction nodes, the phase (free time evolution) is
changed randomly within the possible range of $\langle 0,2
\pi\rangle$. A corresponding physical system could be for example a
network of optical elements with phase fluctuations arising from
different optical lengths between the elements. The fluctuations could
be produced at will with phase shifters to make systematic studies of
the effect of randomness. Alternatively, our calculations can model
undesired phase fluctuations in optical networks and can be used to
study the stability of interference effects in a complicated
interferometer. We show that in the random 2-D network classical
stochastic diffusion is typical.

In the second configuration considered, the network models a 1-D system. Now
the basic unit of the network contains fixed randomness. The basic
unit is formed by the two columns of $A$ and $B$ ("the motif" element
for building the network) in Fig.1, and along the connecting lines
between the columns the phase is adjusted arbitrarily, i.e., within
the range of $\langle 0,2\pi\rangle$. In such a way we define a random
motif matrix. Then this very same basic unit is precisely repeated $M$ times.
The horizontal dimension of the network can now be understood as time
and the network actually models time evolution in one dimension (the
vertical dimension is the real space or a mode space). Note that the
configuration also corresponds to a thin layer of material inside a
cavity, with a light pulse crossing it several times: the 1-D
propagation is now the spreading of the light along the material layer
after many cavity crossings. We show that randomness in the
1-D type configuration leads to quantum localization.

We explain the different behaviour of the two configurations by
relating our observations to the standard theory of localization. The
theory of Anderson localization and dynamical localization is
formulated considering the properties of the Hamiltonian for the
system of interest. This is not always the best possible approach in
the quantum engineering context, where the individual components of a
more complicated network are described by a unitary transform. The
transfer matrix for the total network is easily obtained by simple
multiplications of the transfer matrices corresponding to the
individual elements, but to find out the Hamiltonian corresponding to
the complex system is non-trivial if not impossible \cite{TJ2}.
Therefore we find it of interest to consider here, for the first time
to our knowledge, the quantum localization problem by starting the
theoretical description from the transfer matrices of the system.

Note that in the mathematical level, the unitary transforms describing
the network can also be understood as operations performed in the
computational space of a quantum computer. Thus our investigations
also relate to decoherence affecting certain types of computational
operations. Nearest-neighbour coupling configurations have been
considered in the quantum information context recently
\cite{Wootters00,Briegel01}.

In section 2 we introduce the transfer matrix which is the common
starting point for both network configurations of interest. The first
configuration, leading to classical diffusion, is considered in
section 3, the second, showing localization, in section 4.
Differences in the quantum behaviour in these two cases are explained
with the theory of Anderson/dynamical localization in section 5. The
conclusions are presented in section 6.

\section{The transfer matrix}

The network we consider realizes the nearest-neighbour structure given
in Fig.1. Each element (box in the figure) represents a $2\times 2$
coupler, e.g.\ a beam-splitter in case of optical networks. The whole
network is build up by repeating the motif structure consisting of a
row of components $\fbox{A}$ and $\fbox{B}$, i.e., we use the outputs
of the motif as the input for the next layer. We assume periodic
boundary conditions \cite{Torma98,Kraus}.

The transfer matrix describing the motif unit has the form
\begin{eqnarray}
U = \left[ \begin{array}{cccc}
\fbox{A} & \begin{array}{cc} 0 & 0 \\ 0 & 0 \end{array} &
\begin{array}{c} \cdots \\ \cdots \end{array} & \\
\begin{array}{cc} 0 & 0 \\ 0 & 0 \end{array} & \fbox{A}  & & \\
\begin{array}{cc} \vdots & \vdots \end{array} & & \ddots & \\
& & & \fbox{A} \end{array} \right]
\left[ \begin{array}{ccccc}
B_{22} & \begin{array}{cc} 0 & 0 \end{array} &
\cdots & & B_{21} \\
\begin{array}{c} 0 \\ 0 \end{array} & \fbox{B}  & & & \\
\vdots & & \fbox{B} & & \\
& & & \ddots & \\
B_{12} & & & & B_{11} \end{array} \right] ,
\end{eqnarray}
where
\begin{eqnarray}
\fbox{A} = \left[ \begin{array}{cc}
\cos \theta & \sin \theta \\ - \sin \theta & \cos \theta
\end{array} \right] &;& \fbox{B}=
\left[ \begin{array}{cc}
\cos \phi & \sin \phi \\ - \sin \phi & \cos \phi
\end{array} \right]  . \label{AandB}
\end{eqnarray}
When the parameters $\theta$ and $\phi$ are real, (\ref{AandB})
describes a SU(2) network. Purely imaginary $\theta$ and
$\phi$ correspond to a SU(1,1) network. Above, $\sin^2 x$ ($x$ is
$\theta$ or $\phi$) gives the coupling strengths of the elementary
units used and $N$ is the number of the units in one layer. We will
consider only the case $\cos^2 x = 0.5$. In optics, this corresponds
to so called balanced beam splitters (or 3-dB couplers).

The transfer matrix $W$ of the whole network, in the case all motif
sturctures are the same, is given as a power of $U$:
\begin{equation}
W(M) = U^M .
\end{equation}
The other possibility is that all the motif structures $U_i$,
$i\in [1,M]$ are different due to overall randomness, then
\begin{equation}
W(M) = U_1 D_1 U_2 D_2 ... U_M.
\end{equation}
The matrices $D_i$ represent diagonal matrices giving the phase uncertainities between
the random motifs. Here $M$ can be understood as the size ($2 N\times M$) of
the network in the case of a 2-D network, or as time when the network models
1-D propagation.

An example of the typical interference behaviour in complete absence of
randomness is presented in Fig.2 

\vskip1truecm 
\hskip2truecm\includegraphics[width=5cm,height=5cm]{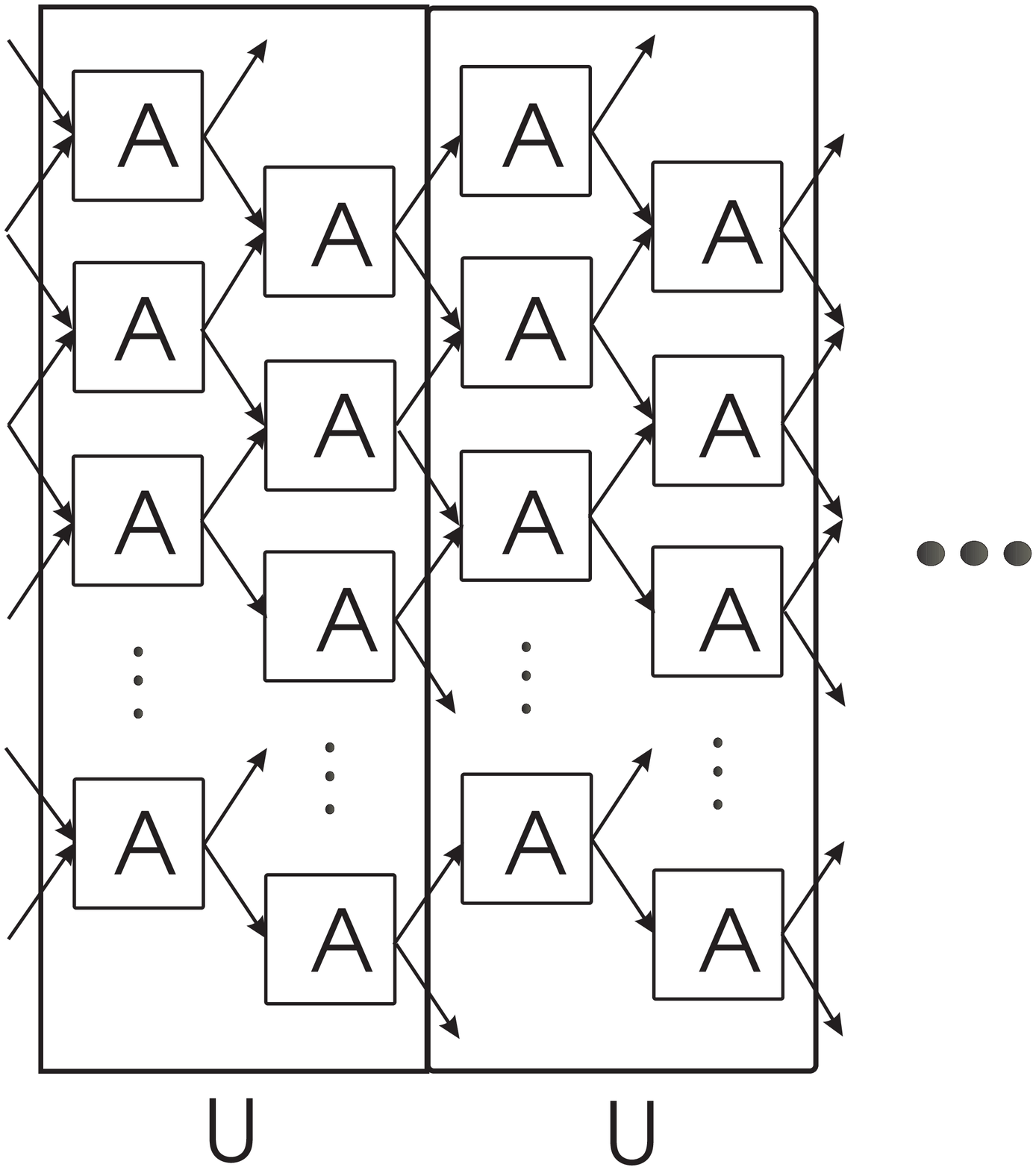}\hskip1truecm\includegraphics[width=5cm,height=5cm,angle=90]{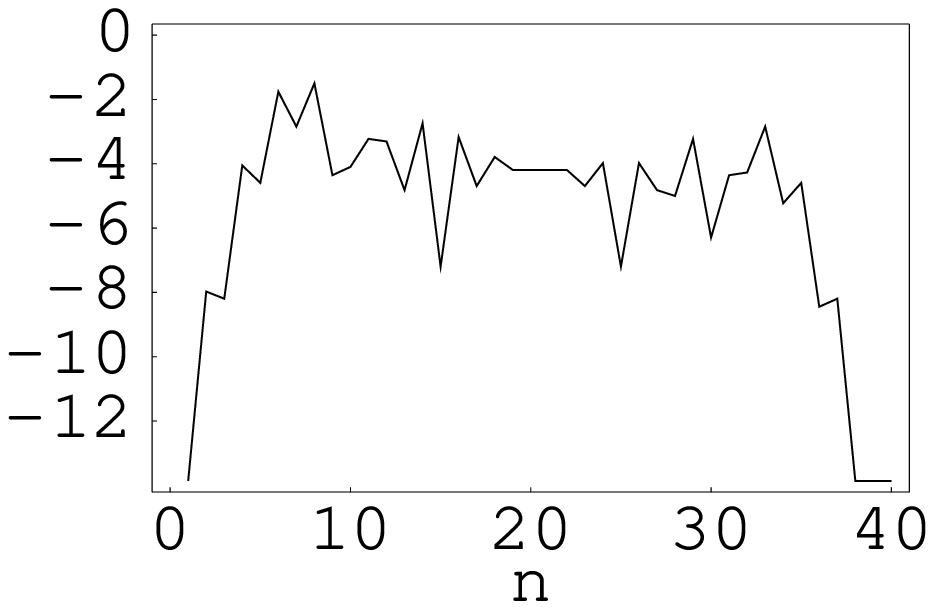}\\

\begin{figure}
\caption{A typical interference pattern in the case of no randomness in
the network. The left picture shows the repetition of the motif structure of the network, the right the logarithm
of the probability $P$ to find the photon at a particular output $n$. We used 40 inputs and outputs. 
The vertical axis shows the labels of the outputs, the horizontal one the logarithm of the probability.}
\end{figure}

In the following we will show how this interference pattern is modified by different forms of randomness.

\section{Random 2-D network}

We consider first a two-dimensional quantum network with
randomness. This corresponds to the total transfer matrix
\begin{equation}
W(M) = U_1 D_1 U_2 D_2 ... U_M ,
\end{equation}
where $U_i, D_i$ are all different. The randomness enters the motif
stuctures either via the individual quantum components $A$
and $B$, or by having an extra random phase shift matrix
between the matrices containing the $N$ components $A$ and $B$.
For optical systems, the first case would correspond to having
fluctuations in the transmittivities of the beam splitters, the latter
to random phase shifts occuring in the optical paths.
Here we have analyzed phase
randomness. The phases between all the connecting points change
randomly within the $\langle 0,2\pi\rangle$ interval.

We study the behaviour of the network by following the propagation
of an input state initially located at the input port which is at
the center of the network. To illustrate our results
we fixed the number of network input ports to $40$ which corresponds to
$20$ elements in one layer. As the initial state we consider a single photon
guided into the network from the $20$-th input port
\begin{equation}
\vert \psi (0)\rangle = \vert 0\rangle_1 ... \vert 0\rangle_{19}
\vert 1\rangle_{20} ... \vert 0\rangle_{40} .
\end{equation}
The spread of a photon given by this initial state in the network is shown in
Fig.3 

\vskip1truecm
\hskip2truecm\includegraphics[width=6cm,height=6cm]{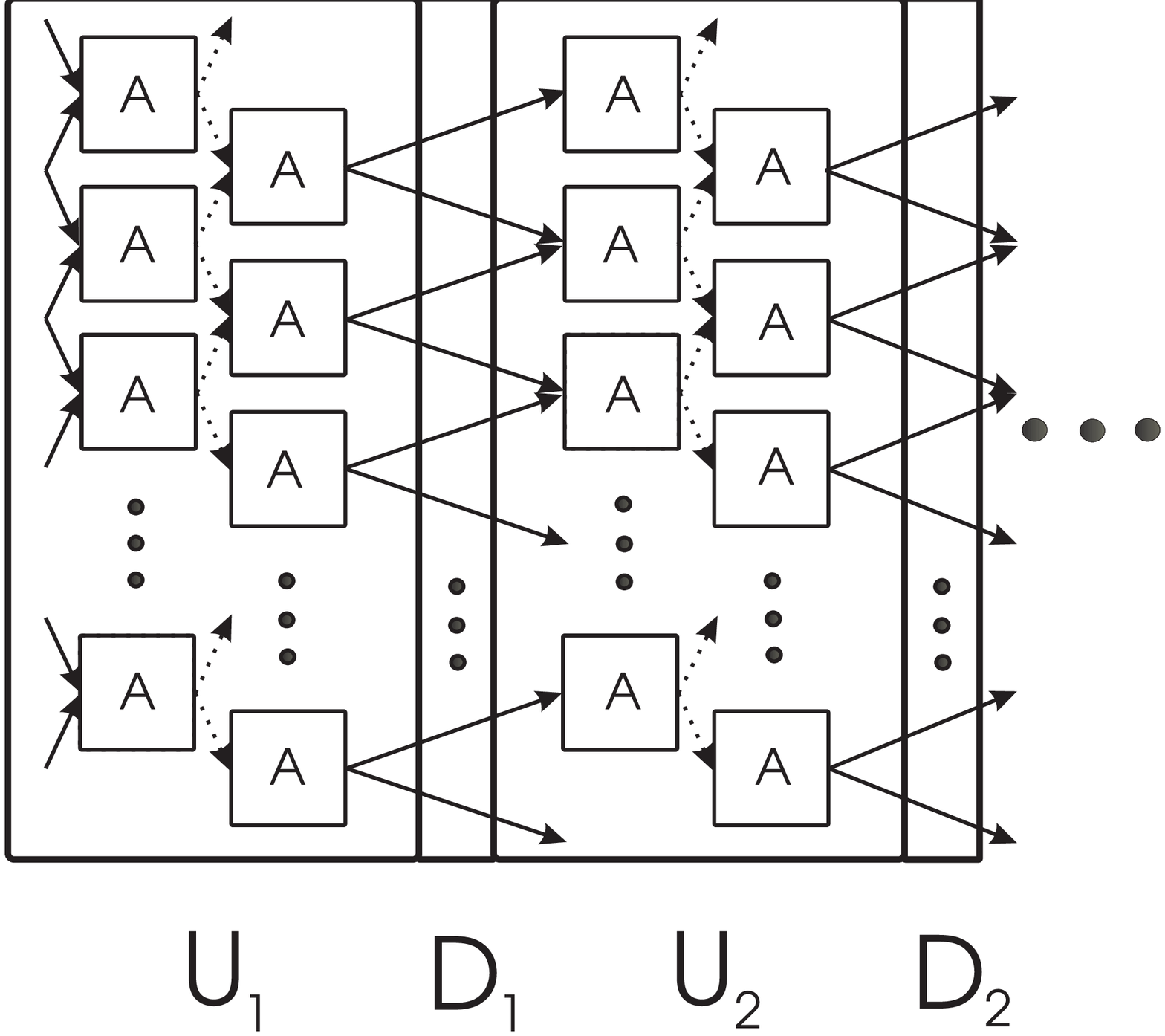}\hskip1truecm\includegraphics[width=5.9cm,height=5.9cm,angle=90]{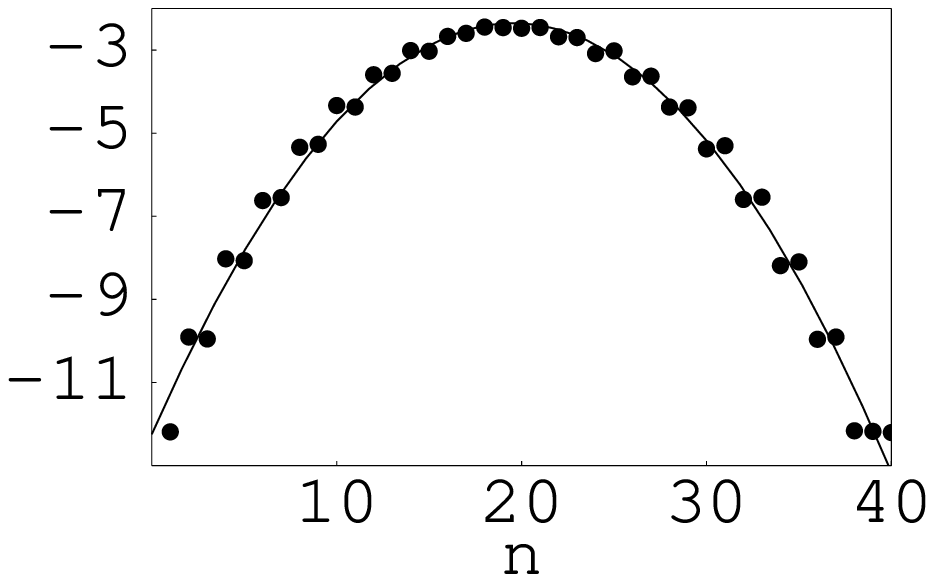}\\ 
\includegraphics[width=15cm,height=4cm]{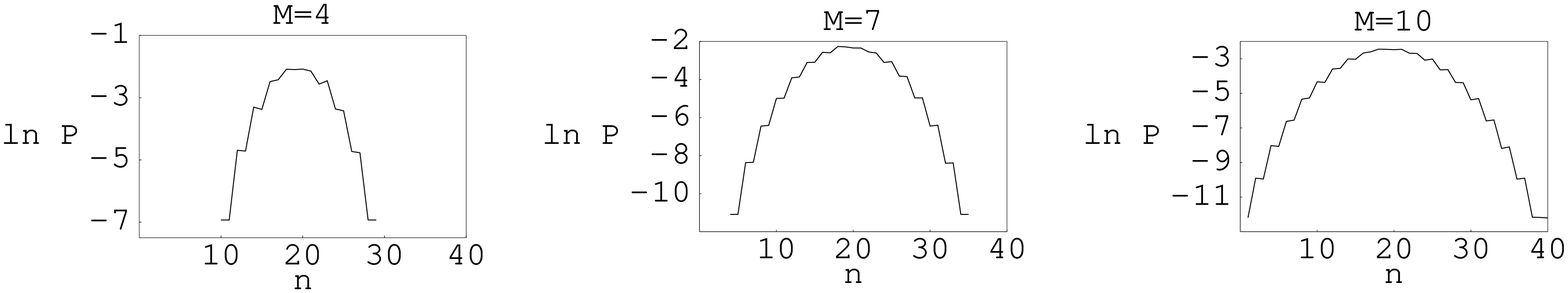}\\

\begin{figure}
\caption{The left picture shows the considered network structure. All $U_i$ are random because
of random phases introduced inside each motif structure, in the figure these phases are 
represented by the dotted arrows connecting the two columns of the elements $A$. Additional random phases are introduced in the layers $D_i$ between each two motifs. 
The output distribution
after ten layers is shown on the right. The dots represent the actual output probabilities, 
the solid line shows the least square fit. The bottom plot shows the
evolution of the distribution after four, seven and ten layers traversed. Note the
change in the logarithmic (vertical) scale.} 
\end{figure}

The plot shows the logarithm of the probability
for the photon to appear at the given output (given by the dots).
Three choices for $M$ are shown to illustrate how the distribution changes
when the photon propagates in the network. The largest value of $M$ equals ten; after
ten layers the photon hits the border of the network, then traversing just another motif
layer the photon would be recycled back into the system due to periodic boundary conditions.
In calculating the probability we have averaged over $1000$ runs.
The solid line gives the least square fit to the calculated values to show that the
distribution follows a quadratic dependence.

As shown in Fig.3, the quantum state in this network behaves like a
classical distribution, i.e.\ it spreads into a broad Gaussian at the
outputs. Classical and quantum behaviour in this kind of networks, in
the absence of randomness, is studied in
\cite{TJ1,Bouwmeester,Harmin,others}. Gaussian outputs were found in
the classical case \cite{Harmin} whereas complicated interference and
recurrence phenomena are typical in the quantum case.  Our results
show that introducing randomness acts as decoherence which reduces the
quantum behaviour to the classical one. In section 5 we explain the
results further in the light of random matrix theory of localization.

\section{Propagation in 1-D; a thin layer of material}

Next we consider the situation where the transfer matrix of
one motif structure of the network has the form $UD$, where
$D$ is a diagonal matrix with $\it fixed$ randomness, and
the total transfer matrix is simply
\begin{equation}
W(M) = (U D)^M .
\end{equation}
Physically this corresponds to quantum propagation in one dimension,
or for example a thin layer of material inside a cavity through which
light makes several passages. The situation is illustrated in Fig.4

\vskip1truecm
\hskip2truecm\includegraphics[width=6cm,height=6cm]{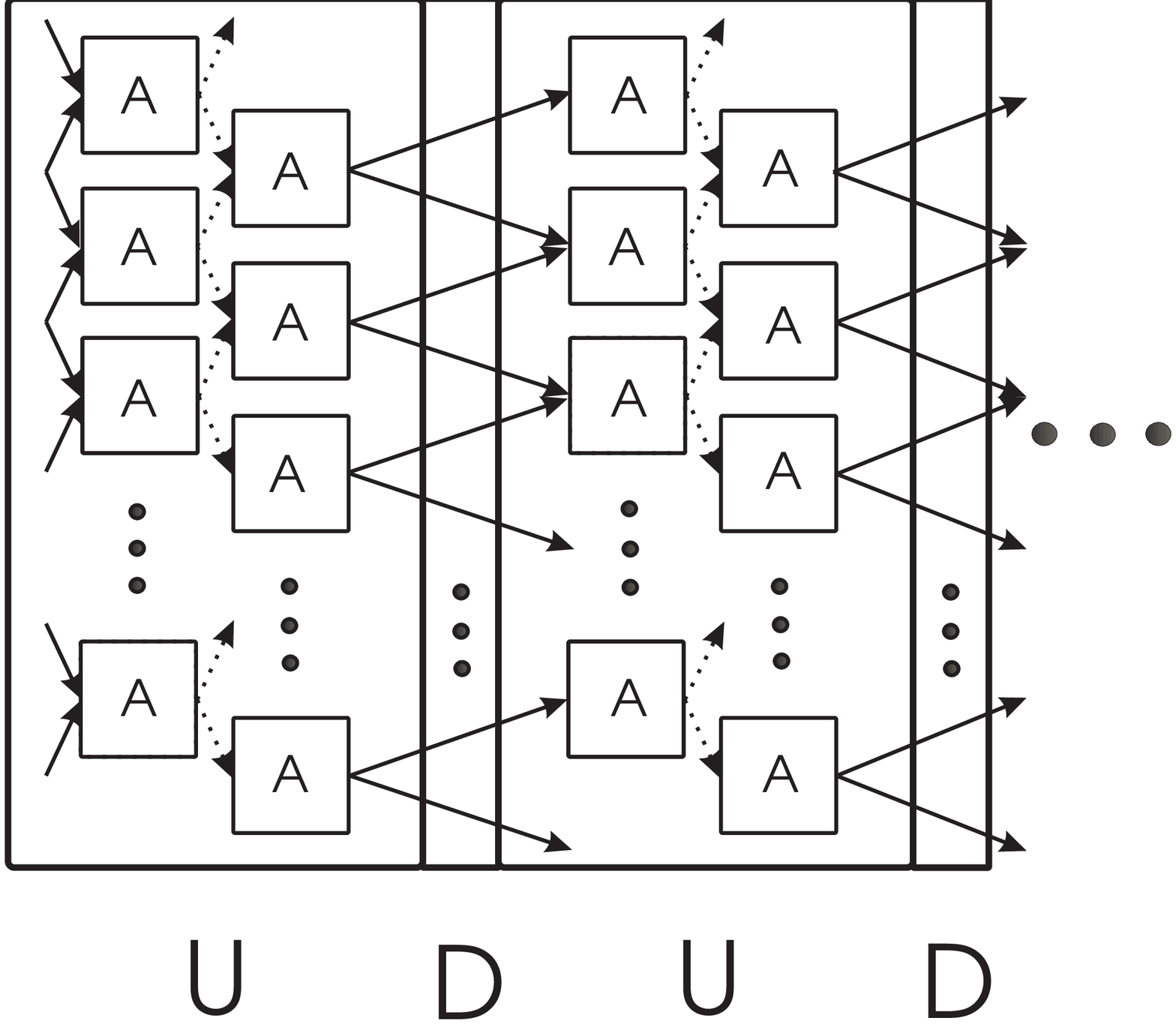}\hskip1truecm\includegraphics[width=5.9cm,height=5.9cm,angle=90]{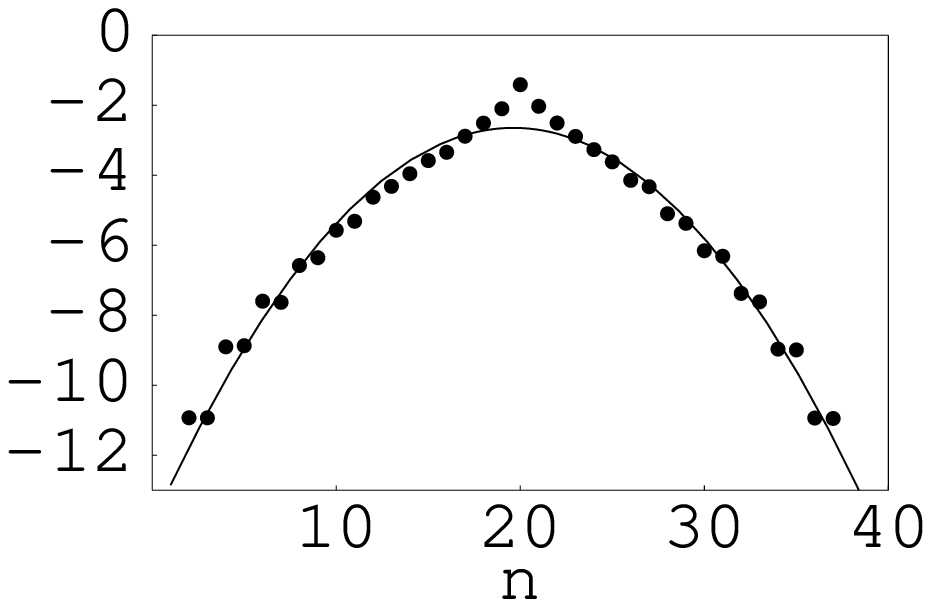}\\ 

\includegraphics[width=15cm,height=20cm]{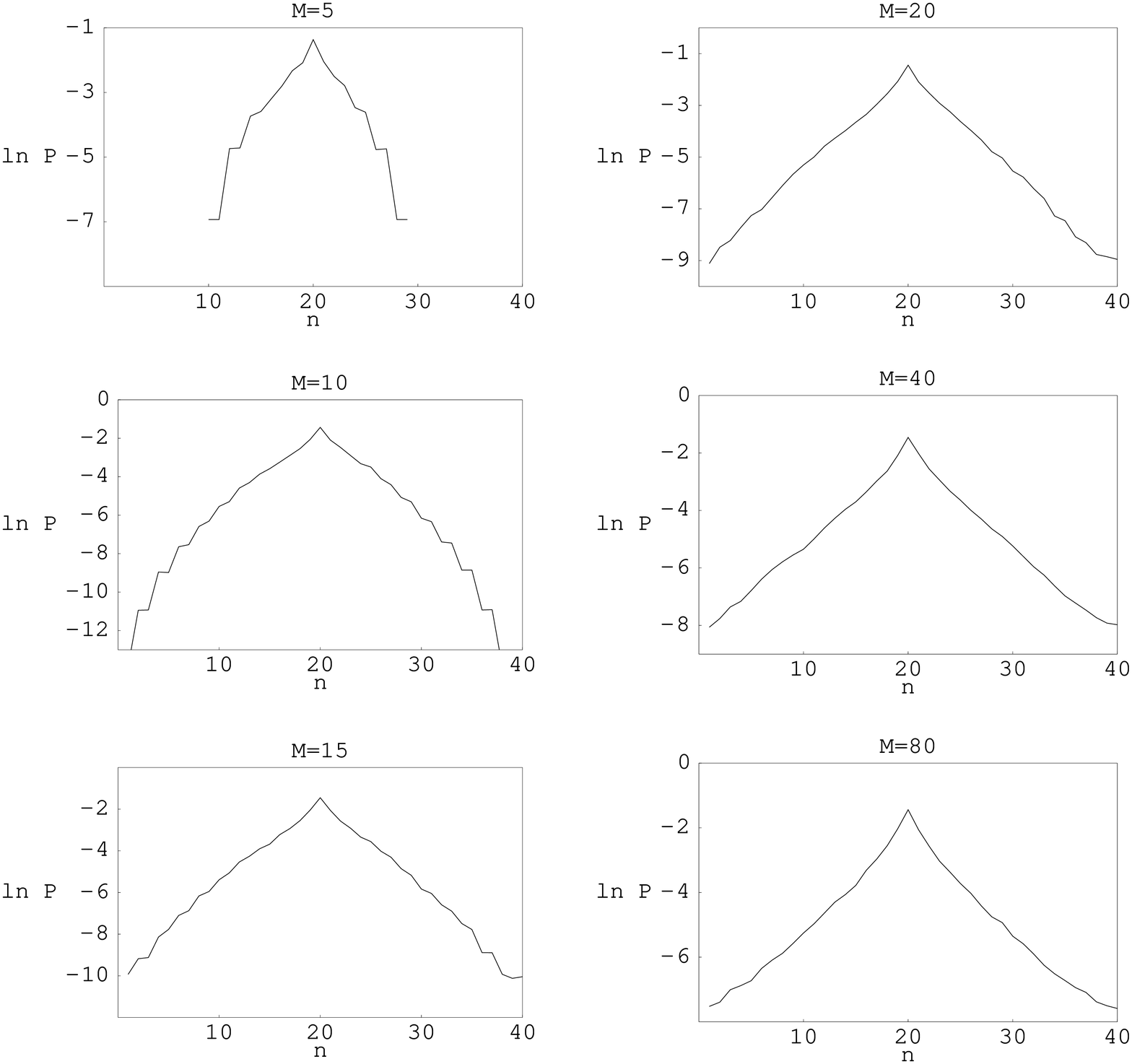}\\

\begin{figure}
\caption{The evolution of the output probability distribution for a network configuration
formed by repeating a fixed random motif. The top right plot shows the distribution after
10 traversed motifs (dots), compared to the corresponding least square fit. The evolution for
different numbers of motifs is given by the three lower pairs of plots. They are for
$M=5,10$, $M=15,20$ and $M=40,80$.}
\end{figure}

We consider the behaviour of the same input as given in the previous section.
Fig.4 shows that the distribution becomes exponentialy peaked at the
input position $20$. To see the emergence of this exponential localization
we present the evolution of the initial photon distribution in dependence on the
number of motifs traversed ($M$).

We have found that the complicated interference pattern typical for
the ideal network turns gradually into quantum localization when the
amount of randomness is increased. This is demonstrated in Fig.5.

\includegraphics[width=15cm,height=20cm]{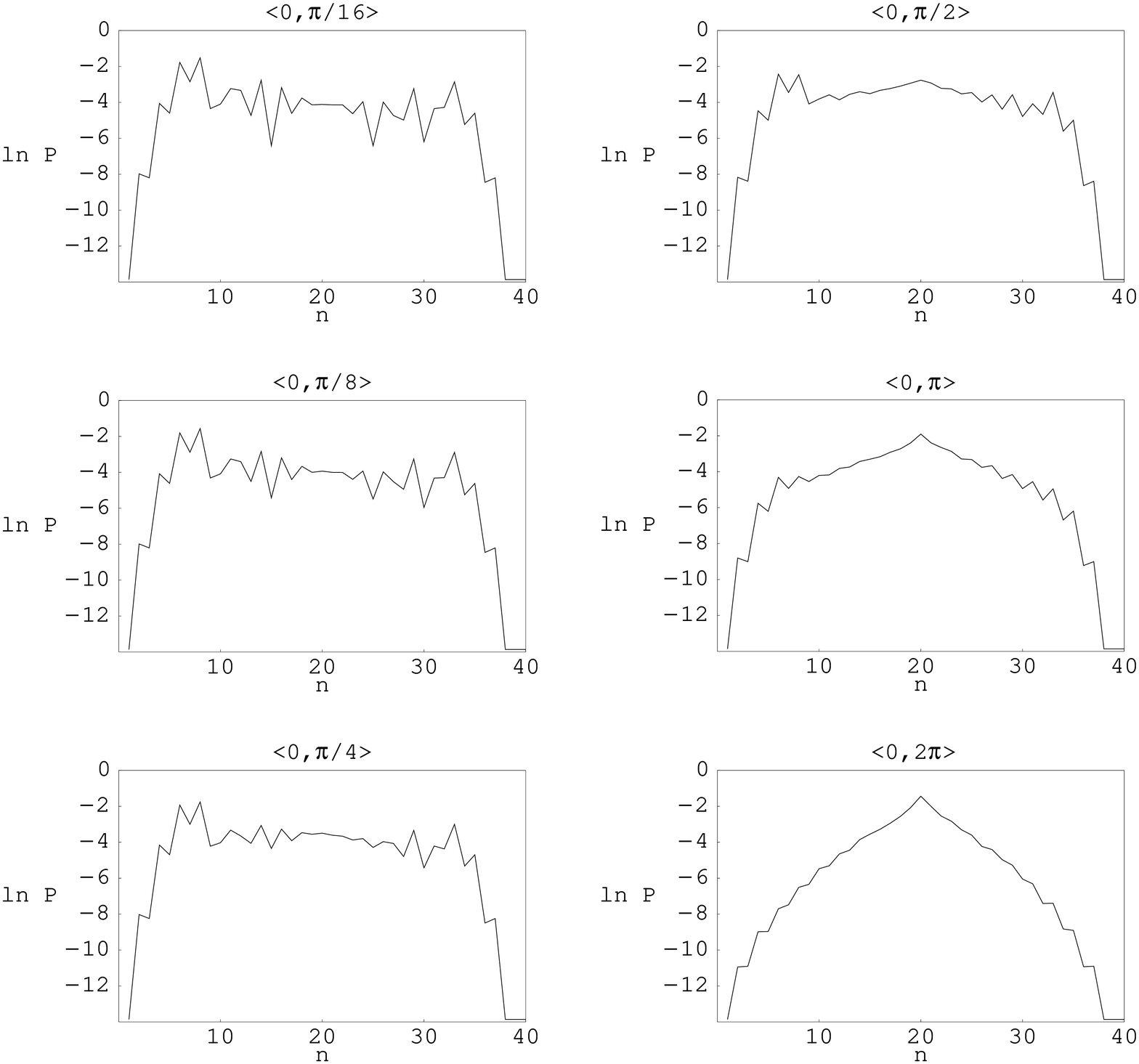}\\

\begin{figure}
\caption{The emergence of localization when the amount of randomness 
in the network is increased. The number of motifs is $M=10$. The interval of
random phases changes from top to bottom and left to right as 
$\langle 0, \pi/16\rangle,\langle0, \pi/8\rangle,\langle 0, \pi/4\rangle,
\langle 0, \pi/2\rangle,\langle 0, \pi\rangle, \langle0, 2\pi\rangle$.
The upper left picture shows the almost unperturbed interference pattern
appearing for very weak randomness, the lower right picture the localized 
distribution in the case of strong randomness.}
\end{figure}

We compare the distributions at
$M=10$ for weak and strong randomness. The strength of randomness
is controlled by the extend to which the phases in the random diagonal
matrices can change, for instance for weak randomness only within the
interval $\langle 0, \pi/16 \rangle$ and for moderate randomness the
interval would be larger $\langle 0, \pi/4\rangle$.
The main result is that we observe exponential localization for strong
enough randomness.  This remarkable quantum feature will be explained
in section 5 by making a connection to the standard theory
of Anderson and dynamical localization.

\subsection{Intermediate case: extra noise in 1-D propagation or
semi-random 2-D network}

As an intermediate case we consider a network with the transfer matrix
\begin{equation}
W(M) = U D D_1 U D D_2 U D D_3 ... ,
\end{equation}
where $D_i$ are random and different from each other. The matrix $D$ is
also random. Note that with all $D_i= 1$, $W(M)$ would be the same
as in the previous section, that is, expected to lead to localization.
This intermediate case models a network with phase
fluctuations inbetween exactly similar motif structures.
Examples of this are an optical network where the nearest neighbour
coupling could be essentially perfect but the path lengths inbetween two
nearest neighbour coupling layers fluctuate, or fluctuating cavity
length in the 1-D case of a layer of material in a cavity.
In modelling 1-D propagation, this is also a test of numerical
stability of the localization phenomenon when very small extra randomness
is introduced.

We now fix the amount of randomness in the matrix $D$ that appears
repeatedly in $W(M)$ (phases chosen between $\langle 0,2\pi\rangle$), and
vary the amount of randomness in $D_i$.
We observe that for weak randomness in $D_i$
(phases chosen between $\langle 0, 0.1 \pi\rangle$),
the localization is preserved (note that
in this case we probably have approximately the situation $W \sim (U D)^M$).
For strong randomness (phases chosen between $\langle 0, 2\pi\rangle$)
Gaussian diffusion is observed. This transition from localization
into diffusion is illustrated in Fig.6.

\vskip1truecm
\includegraphics[width=5cm,height=4cm]{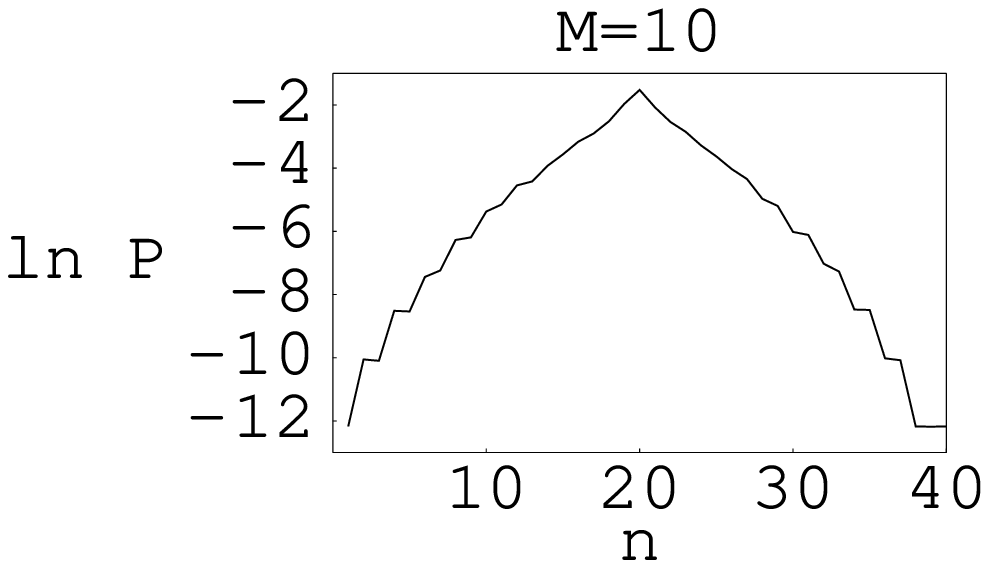}
\includegraphics[width=5cm,height=4cm]{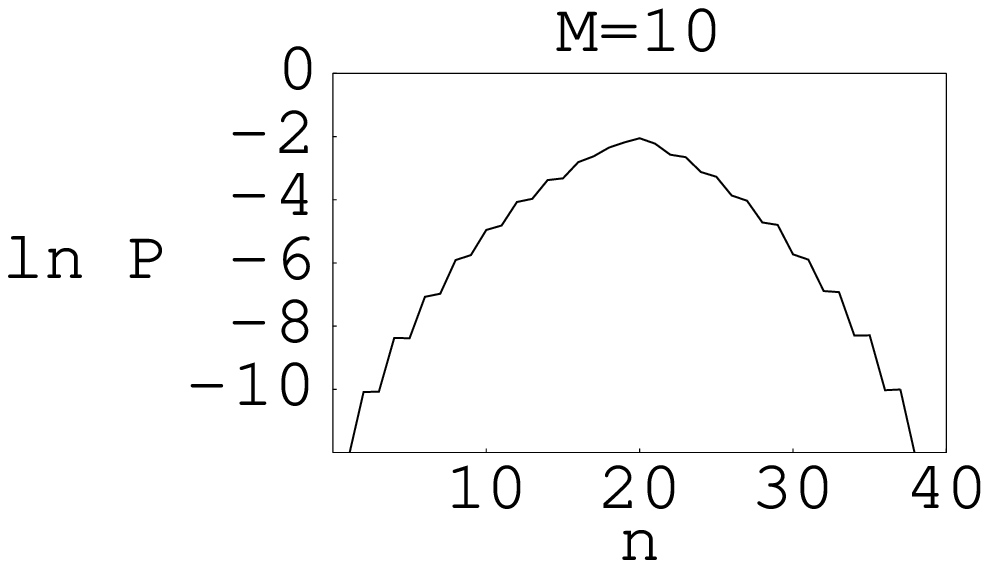}
\includegraphics[width=5cm,height=4cm]{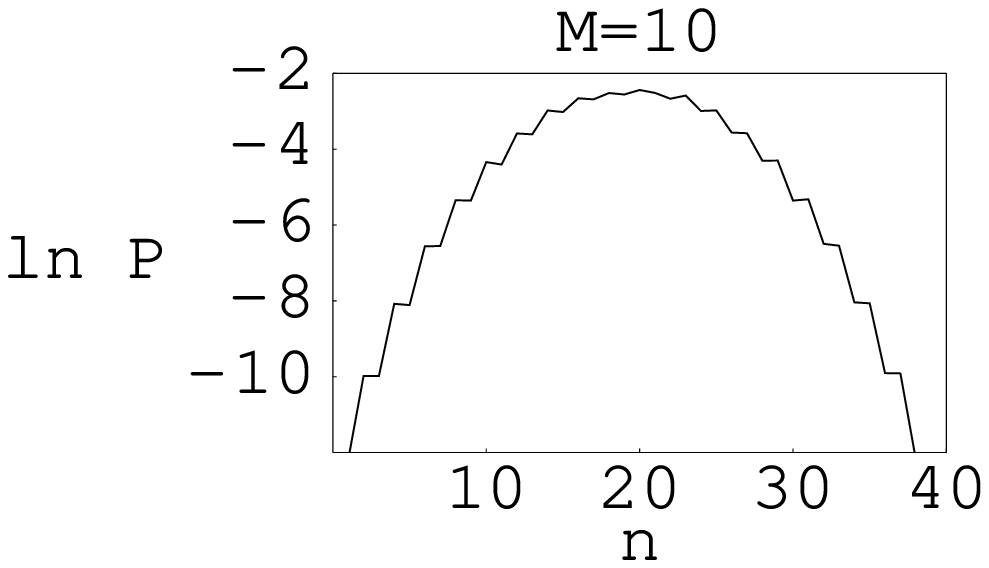}\\

\includegraphics[width=5cm,height=4cm]{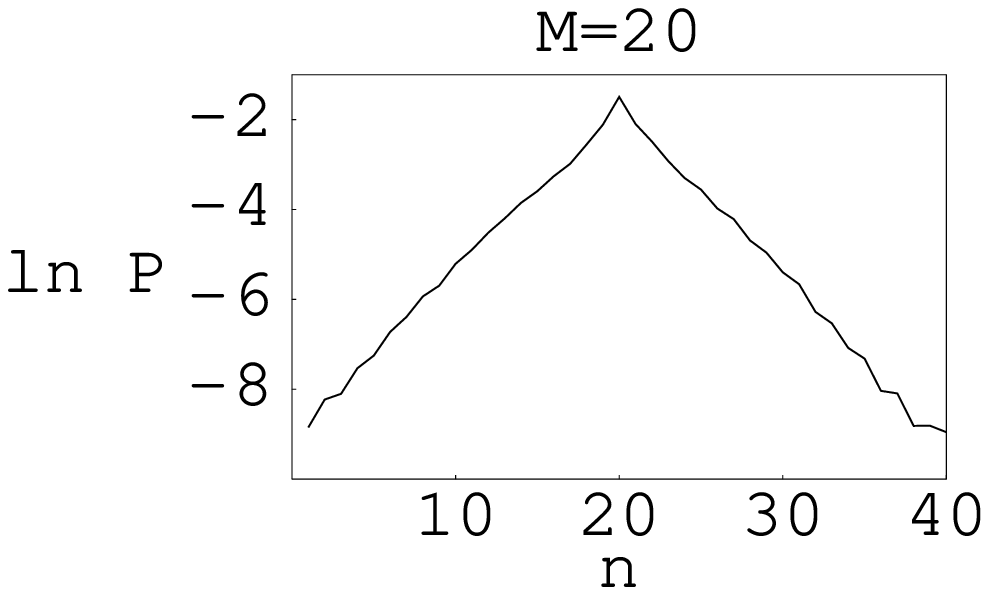}
\includegraphics[width=5cm,height=4cm]{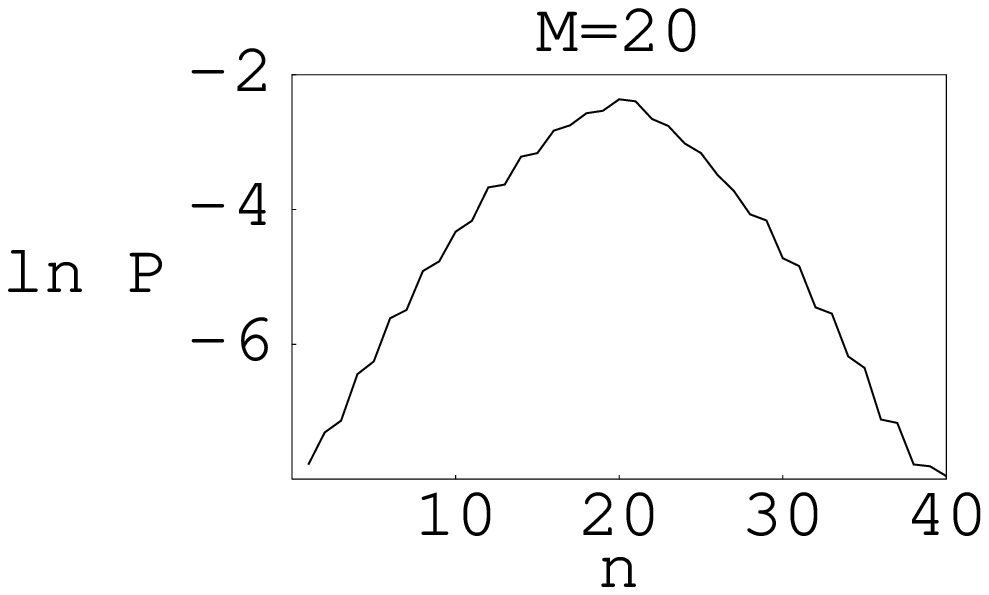}
\includegraphics[width=5cm,height=4cm]{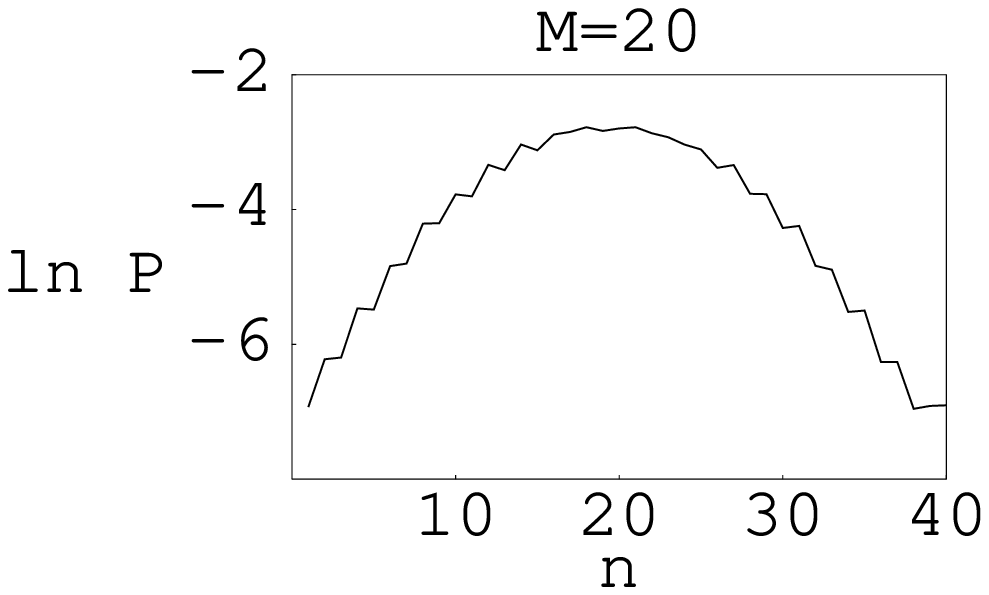}
\\

\begin{figure}
\caption{The transition from localization into diffusion. 
The number of traversed motifs is $M=10$
(top) and $M=20$ (bottom). The left column represents weak extra randomization 
(interval of the $D_i$ phases only $\langle 0,0.1\pi\rangle$). The middle is
an intermediate regime with phases varying $\langle 0,0.1\pi\rangle$. The right column
was calculated for phases randomly chosen from the interval $\langle 0,2\pi\rangle$.
In this case no localization is observed.}
\end{figure}

\section{Localization --- Hamiltonian versus transfer matrix approach}

The theory of Anderson and dynamical localization tells that
exponential localization of the system eigenvectors, and consequently, of
quantum wave packets, occurs when the system Hamiltonian is a random
band-diagonal matrix \cite{Haake92}. Band-diagonality means that only
a few neighbouring sites (thinking in terms of a lattice) are coupled
--- in the Lloyd model \cite{Haake92}, which is the simplest model showing
localization, only the nearest neighbours are connected. Randomness
usually means that the energies of the sites, that is, the diagonal
terms of the Hamiltonian, are randomly chosen according to a certain
distribution. But randomness can equally well appear in the
off-diagonal couplings.

We describe the network of interest by a transfer matrix rather
than a Hamiltonian. Connection between these two is non-trivial to
obtain for large networks, c.f \cite{TJ2}. But some overall characteristics
are easy to sketch. One motif structure of our network clearly couples
nearest neighbours, which corresponds to a Hamiltonian $H_{motif}$
with only nearest neighbour coupling. In the case of 1-D propagation,
or a thin layer of material in a cavity, the total transfer matrix is
\begin{equation}
W(M) = U^M = e^{iH_{motif}M} .
\end{equation}
The network size $M$ can be interpreted as time, and the Hamiltonian
of the system is the one describing one motif structure: random and
band diagonal. Therefore it is in accordance with the theory of
localization to observe an exponentially localized distribution at the
output. The case with overall randomness in a 2-D network is completely
different. The total transfer matrix
\begin{equation}
W(M) = U_1 D_1 U_2 D_2 ... U_M
\end{equation}
can no longer be expressed as an $M$th power of a certain
band-diagonal Hamiltonian.
By diagonalizing $W(M)$ one can obtain the corresponding Hamiltonian,
but it is clear that it will not any more be band-diagonal, but
instead couples $M$ neighbours. For large $M$, the Hamiltonian approaches
a full random matrix. It is known that for full random matrices there
is no localization \cite{Haake92}. Instead, we observe classical-type diffusion.

\section{Conclusions}

We have considered randomness and localization in a specific quantum
network which is characterized by nearest-neighbour structure and has
a relation to models of statistical mechanics with nearest-neighbour
interaction. The physical systems that can be modelled by such a
network are for instance beam splitter networks \cite{Zeilinger},
micro- and nanostructures and networks of intersecting energy levels
\cite{others,SD}. We showed that in a 2-D network with overall
randomness, the quantum behaviour was reduced to the classical one. In
contrast, in 1-D propagation, or a system like a layer of material in
a cavity, we observe exponential localization. An intermediate case
was showing a smooth transition between these two completely different
behaviours.

We relate our findings
to the standard theory of localization. Our discussion emphasises
the new features which arise when the systems of interest are described
by transfer matrices rather than Hamiltonians. This would be the
natural approach in case of engineered quantum networks.

{\it Acknowledgements}

P.T. acknowledges the support by the Academy of Finland (projects
42588, 47140 and 48845 and Research Centre for Computational Science
and Engineering, project 44892, Finnish Centre of Excellence Programme
2000-2005).  The financial support by the Alexander von Humboldt
foundation, MSM 210000018 (Czech Republic), GACR 202/01/0318 and EU IST-1999-13021 for I.J. is
greatefully acknowledged.

\end{document}